\documentclass[twocolumn]{aastex63}

\newcommand\be{\begin{eqnarray}}
\newcommand\ee{\end{eqnarray}}
\usepackage{amsmath}
\usepackage{subfigure}
\usepackage{graphicx}
\usepackage{gensymb}
\usepackage{CJK}

\shorttitle{LAMOST WDs with infrared excess}
\shortauthors{Wang et al.}
\begin{document}
\begin{CJK*}{UTF8}{gbsn}

\title{White Dwarfs with Infrared Excess from LAMOST Data Release 5}
\correspondingauthor{Xiaoxia Zhang; Junfeng Wang}
\email{zhangxx@xmu.edu.cn; jfwang@xmu.edu.cn}
\author{Lin Wang(汪琳)}
\affiliation{Department of Astronomy, Xiamen University, Xiamen, Fujian 361005, China}
\author[0000-0003-4832-9422]{Xiaoxia Zhang (张小霞)}
\affiliation{Department of Astronomy, Xiamen University, Xiamen, Fujian 361005, China}
\author[0000-0003-4874-0369]{Junfeng Wang (王俊峰)}
\affiliation{Department of Astronomy, Xiamen University,  Xiamen, Fujian 361005, China}
\author[0000-0002-2419-6875]{Zhi-Xiang Zhang (张志翔)}
\affiliation{Department of Astronomy, Xiamen University,  Xiamen, Fujian 361005, China}
\author[0000-0002-2853-3808]{Taotao Fang (方陶陶)}
\affiliation{Department of Astronomy, Xiamen University,  Xiamen, Fujian 361005, China}
\author[0000-0003-3137-1851]{Wei-Min Gu (顾为民)}
\affiliation{Department of Astronomy, Xiamen University,  Xiamen, Fujian 361005, China}
\author[0000-0002-8321-1676]{Jincheng Guo (郭金承)}
\affiliation{Department of Scientific research, Beijing Planetarium, Xizhimenwai Road, Beijing 100044, China}
\author[0000-0002-3878-5590]{Xiaochuan Jiang (姜小川)}
\affiliation{Department of Astronomy, Xiamen University,  Xiamen, Fujian 361005, China}

\begin{abstract}

Infrared excess is an important probe of sub-stellar companions and/or debris disks around white dwarfs (WDs). Such systems are still rare for in-depth understanding of their formation and long-term evolution. One of the largest spectroscopic surveys carried out by the Large sky Area Multi-Object fiber Spectroscopic Telescope (LAMOST) recently released more than $3000$ WDs, a significant fraction of which have not undergone excess search. Here we present cross-correlation of LAMOST DR5 WD catalog with the Pan-STARRS, SDSS, UKIDSS, 2MASS, and {\it WISE}. By performing SED (spectral energy distribution) fitting for 846 WDs with $WISE$ detections, we identify 50 candidates with infrared excess, including 7 candidate WD+M dwarf binaries, 31 candidate WD+brown dwarf (BD) binaries and 12 candidate WD+dust disk systems. 8 of the dust disk systems are our new identifications. Utilizing a systematic survey with accurate stellar parameters derived from spectral fitting, our work is an important addition to previous searches for infrared excess from SDSS and {\it Gaia} WDs, and provides a significant ($\gtrsim8\%$) complement to current database of WDs with candidate BD companions and dust disks. The frequencies of WD+BD binaries and WD+dust disk systems are constrained to be $\lesssim3.7\%$ and $\sim1.4\%$, respectively. The properties of candidate dust disk systems are discussed.  All of our candidates require follow-up observations for confirmation owing to limited spatial resolution of {\it WISE}.
\end{abstract}


%%%%%%%%%%%%%%%%%%%%%%%%%%%%%%%%%%%%%%%%%%%%%%%%%%%%%%%%%
\section{Introduction} 
\label{sec:intro}

White dwarfs (WDs) are the final evolutionary stage and remnants of most stars \citep[$\lesssim 8-10 M_\odot$;][]{1997ApJ...489..772I}.  
Infrared excess around a WD may come from an M dwarf companion, a brown dwarf (BD) companion, or a debris disk surrounding it.  A statistical sample of WD+BD binaries and/or WD+debris disk systems are crucial for testing long-term evolutionary models of planetary systems around stars of $\lesssim 8-10 M_\odot$.

BDs are substellar objects of $\sim13-80$ Jupiter masses, too light to sustain nuclear fusion of hydrogen but sufficient for deuterium burning \citep[e.g.,][]{2000ARA&A..38..337C, 2011ApJ...736...47B}. Observations of BD companions to main-sequence stars have revealed a paucity of separation within $\sim 3$ au, namely ``BD desert" \citep[e.g.,][]{2000PASP..112..137M, 2006ApJ...640.1051G, 2016AJ....151...85T}. Various explanations have been proposed, including orbital migration \citep[e.g.,][]{2002MNRAS.330L..11A}, tidal interactions with the host star \citep[e.g.,][]{2002ApJ...568L.117P, 2016A&A...589A..55D}, and different formation mechanisms for low- and high-mass BD companions \citep[e.g.,][]{2014MNRAS.439.2781M}. The formation of BDs itself is still an open question. Possible scenarios include a scaled-down version of star formation process \citep[star-like;][]{2003MNRAS.339..577B}, gravitational instabilities in the protoplanetary disk \citep[planet-like;][]{2014prpl.conf..619C}, etc. 
Detached WD$-$BD pairs offers a valuable opportunity to investigate the “BD desert” and the formation of BDs, as those BDs have survived the death of the host star via the poorly understood common-envelope phase of close binary evolution \citep{2013A&ARv..21...59I}. The occurrent rate of WD+BD binaries is estimated to be $0.5-2$ percent \citep[e.g.,][]{2011MNRAS.416.2768S, 2011MNRAS.417.1210G}, and to date less than a dozen have been confirmed \citep[e.g.,][]{2018MNRAS.481.5216C}.

Debris disks are believed to form through tidal disruption of asteroids (or comets) perturbed from locations comparable to the Kuiper belt in our solar system \citep[][]{2011MNRAS.414..930B} into the Roche lobe of the WD \citep[$\sim1 R_\odot$;][]{1999Icar..142..525D} by a remnant planetary system \citep[e.g.,][]{2003ApJ...584L..91J, 2012ApJ...747..148D, 2014MNRAS.442L..71V}. Theoretical studies predict that a fraction of planets can survive the red giant phase of the host star \citep[e.g.,][]{2007ApJ...661.1192V, 2009ApJ...705L..81V, 2010MNRAS.408..631N, 2012ApJ...761..121M}, which is supported by discoveries of debris disks \citep[e.g.,][]{2006Sci...314.1908G, 2009ApJ...694..805F, 2016MNRAS.459.1415B}, disintegrating planetary bodies \citep[e.g.,][]{2015Natur.526..546V, 2016ApJ...816L..22X}, and recently planets \citep[][]{2020Natur.585..363V, 2021Natur.598..272B} orbiting around WDs. Dust grains in the disk will spiral in towards the WD via Poynting-Robert drag \citep{2011ApJ...732L...3R}, and the sublimated dust will accrete onto the WD's surface due to viscous torques \citep[][]{2012ApJ...760..123R, 2013MNRAS.431.1686V}. This results in a spectroscopically detectable pollution of the otherwise pristine WD atmosphere, since the gravitational settling time is short compared to the WD's cooling age and the metals will rapidly diffuse out of the WD's photosphere \citep[e.g.,][]{2009A&A...498..517K}. 
Observations suggest that at least $20-30$ percent of WDs are polluted by metals \citep[e.g.,][]{2003ApJ...596..477Z, 2010ApJ...722..725Z, 2014A&A...566A..34K}.

A typical observational feature of debris disks around WDs is an infrared excess over the WD photosphere. The first infrared excess around a WD was detected by \cite{1987Natur.330..138Z} and was later interpreted as emission from circumstellar dust by \citet[][]{1990ApJ...357..216G}. Afterwards, lots of efforts have been put into searches of infrared excess around WDs \citep[e.g.,][]{2011ApJS..197...38D, 2020ApJ...902..127X, 2021ApJ...920..156L}. Currently, about 100 WDs have been identified to host debris disks \citep[e.g.,][]{2013ApJ...770...21H, 2015MNRAS.449..574R, 2015ApJ...810L..17G, 2016NewAR..71....9F, 2020MNRAS.494.2861R, 2021ApJ...920..156L}, largely owing to the sensitive infrared observations provided by {\it Spitzer Space Telescope} \citep[][]{2004ApJS..154....1W}. However, {\it Spitzer} mostly focused its observations on WDs with polluted atmosphere \citep[e.g.,][]{2009ApJ...694..805F, 2012ApJ...745...88X} or in a specific range of stellar effective temperature \citep[e.g.,][]{2015MNRAS.449..574R, 2019MNRAS.487..133W}. The frequency of debris disks around WDs is constrained to be $1-4$ percent, and for heavily polluted WDs, the occurrent rate of dusty disks is found to exceed $50$ percent \citep[e.g.,][]{2007ApJS..171..206M, 2009ApJ...694..805F, 2012ApJ...760...26B, 2015MNRAS.449..574R, 2019MNRAS.487..133W}. 

Infrared surveys such as {\it Wide-field Infrared Survey Explorer} \citep[{\it WISE};][]{2010AJ....140.1868W} and UKIRT Infrared Deep Sky Survey \citep[UKIDSS;][]{2007MNRAS.379.1599L} provide opportunities for comprehensive searches of infrared excess around WDs. The first large and deep untargeted search for infrared excess were carried out by \citet{2011MNRAS.416.2768S} and \cite{2011MNRAS.417.1210G}, via cross-correlating Sloan Digital Sky Survey \citep[SDSS;][]{2000AJ....120.1579Y} WDs with UKIDSS catalog. The release of SDSS DR7 WD catalog \citep{2013ApJS..204....5K} further triggered excess search via cross-correlation with {\it WISE} all-sky survey \citep[][]{2011ApJS..197...38D, 2014ApJ...786...77B}, although source confusion is unavoidable due to the poor spatial resolution ($\sim 6\arcsec$) of {\it WISE} \citep[e.g.,][]{2020ApJ...891...97D}. More recently, {\it Gaia} released its second WD catalog \citep{2019MNRAS.482.4570G}, providing the largest photometric WD sample for infrared excess search via colors and/or magnitudes \citep[][]{2019MNRAS.489.3990R, 2020ApJ...902..127X, 2021ApJ...920..156L}. These studies identified hundreds of candidates with infrared excess in total, and candidate WD+BD binaries and WD+dust disk systems are fewer ($\lesssim100$), since only part of the studies classified them further via detailed spectral energy distribution (SED) modeling.  

The Large sky Area Multi-Object fiber Spectroscopic Telescope (LAMOST) is a powerful spectroscopic survey operated by National Astronomical Observatories, Chinese Academy of Sciences \citep{2012RAA....12.1197C}. In the past ten years, LAMOST has released thousands of spectroscopically identified WDs with stellar parameters (e.g., effective temperature $T_{\rm eff}$, surface gravity $\log g$) derived from spectral fitting \citep{2013AJ....145..169Z, 2013AJ....146...34Z, 2015MNRAS.454.2787G, 2022MNRAS.509.2674G}, which should in principle be more accurate and reliable than that obtained from SED fitting if the signal-to-noise ratio (S/N) of the spectrum is above some criterion, e.g., S/N$>30$ \citep{2022MNRAS.509.2674G}.  The vast majority of LAMOST WDs locate at low Galactic latitudes \citep{2012RAA....12..723Z, 2012RAA....12..735D}, providing a complement to SDSS DR7 WD catalog \citep{2013ApJS..204....5K}. Recently, LAMOST released its 5$th$ spectroscopically identified WD catalog \citep{2022MNRAS.509.2674G}. More than $1000$ of them have not been searched for infrared excess, offering opportunities for discoveries of new targets of interest, e.g., WDs with a debris disk or a BD companion. Without cut-off in the spectral type or effective temperature of the WDs, this catalog provides an unbiased sample in this point of view, although other selection effects may exist (see discussions in Section~\ref{sec:rate}). In this paper, we present systematic search for infrared excess in the LAMOST DR5 WD catalog according to {\it WISE} photometry. The poor spatial resolution of {\it WISE} make background contamination unavoidable, which is a shortcoming of this work.  

The paper is organized as follows. In Section~\ref{sec:data}, we present photometric data of our sample by cross-matching with various surveys in optical and infrared. Then we describe SED models of different components and search for infrared excess in Section~\ref{sec:model}. We further check the optical and infrard images of our candidates in Section~\ref{sec:image} to exclude possible background contamination or M dwarf companions that are not the focus of this work. Results and discussion are presented in Section~\ref{sec:result} and conclusions are given in Section~\ref{sec:conclusion}.

\section{Photometric data}
\label{sec:data}

LAMOST DR5 WD catalog contains 3064 unique sources \citep[see][Table 3]{2022MNRAS.509.2674G}.
To search for infrared excess around those WDs via SED fitting in Section~\ref{sec:model}, prior information on stellar parameters (e.g., $T_{\rm eff}$ and $\log g$) is required to simultaneously model the radiation of the WD photosphere as well as an additional component (if any) with the lack of data at wavelengths longer than several mirons. Moreover, $Gaia$ distance is adopted to convert the absolute magnitude to radiative flux for various photometric systems when applying WD cooling model  \citep{1995PASP..107.1047B}. For these two reasons, we only select targets with $T_{\rm eff}$, $\log g$, and $Gaia$ distance measurements. Some of the targets have multiple observations, and each spectrum has best-fit values of $T_{\rm eff}$ and $\log g$. We adopt the fitting parameters derived from the highest S/N spectrum. This results in a sample of 1179 WDs (hereafter the parent sample), and all of them were classified as either DA or DB types\footnote{The stellar parameters of LAMOST WDs were fitted only for DA and DB types by \cite{2022MNRAS.509.2674G}. For other spectral types, the modeling is more complicated and the spectral fitting was not performed.} \citep{2022MNRAS.509.2674G}.
 
Then we search for optical and infrared data by cross-correlating the LAMOST coordinates of the parent sample with various surveys within a radius of $2\arcsec$ \citep[e.g.,][]{2011ApJS..197...38D}. Our excess search mainly relies on {\it WISE} photometry, with $W1$, $W2$, $W3$, and $W4$ bands centered at $3.4$, $4.6$, $12$,  and $22$ \micron, respectively. 
Cross-correlating with AllWISE catalog results in $459$, $303$, $15$ and $7$ targets, respectively for $W1-W4$ detections. CatWISE2020 catalog \citep{2021ApJS..253....8M} provides an excellent alternative for $W1$ and $W2$ bands as the detection limit is fainter and the sample size is roughly twice larger than AllWISE. There are $828$ and $685$ of the parent sample having CatWISE2020 detections in $W1$ and $W2$ bands, respectively.  For $W1$ and $W2$ photometry, we give priority to CatWISE2020 data, and if is not available, we adopt AllWISE data (the case for 18 targets). Since $W3$ and $W4$ data are not provided by CatWISE2020, we directly use AllWISE data if available. 

We then cross-match the parent sample with the Two-Micron All Sky Survey \citep[2MASS;  $JHK_{s}$;][]{2006AJ....131.1163S}. Considering the limited resolution of 2MASS, we further cross-match these targets with UKIDSS catalog from Galactic Plane, Galaxy Cluster, and Large Area Survey \citep[{\it YJHK};][]{2007MNRAS.379.1599L}.  
$293$ and $292$ of the parent sample are detected in the 2MASS $K_s$ and UKIDSS $K$ bands, respectively, and $77$ of the targets are detected by both filters. 
We notice that some of the WDs have inconsistent photometry for 2MASS and UKIDSS data. Since they have similar wavelength coverage and UKIDSS has better spatial resolution, we give priority to UKIDSS data and adopt 2MASS data only if the corresponding band UKIDSS data are not available.  

In order to characterize the photospheric radiation of WDs, the parent sample are further cross-matched with Pan-STARRS \citep[$g_p\ r_p\ i_p\ z_p\ y_p$;][]{2010SPIE.7733E..0EK} catalog. As a consistency check, we also search for their SDSS DR12 detections \citep[{\it ugriz};][]{ 2015ApJS..219...12A}. This returns $1110$ and $895$ detections for $g_p$ and $g$ bands of the two surveys, respectively, and most of these targets have the other four optical-bands observations.

Except for the above-mentioned cases of either CatWISE2020 or AllWISE, and either 2MASS or UKIDSS, all of these optical and infrared data are used in the SED fitting if the target is detected in that filter, i.e., the apparent magnitude is given with errors instead of upper limits. We add additional 2\% in quadrature to the reported uncertainties in the radiative flux in each photometric band to account for systematic errors.

\section{Models and SED fitting}
\label{sec:model}

In order to search for infrared excess around the WDs and distinguish among possible scenarios for their origin, we adopt $T_{\rm eff}$ and $\log g$ measurements of LAMOST WD catalog derived from spectral fitting of Balmer absorption lines \citep{2022MNRAS.509.2674G} in the SED fitting, and our models involve (i) photospheric emission of a WD; (ii) the SED of a late-type (M/brown dwarfs) companion; and (iii) radiation spectrum of a dust disk, as detailed below.
\begin{itemize}
\item[(i)] Photospheric emission of WDs. For a grid of $T_{\rm eff}$ and $\log g$, WD cooling models\footnote{The cooling model data are publicly available online: https://www.astro.umontreal.ca/$\sim$bergeron/CoolingModels/. The model results are given for $T_{\rm eff}$ in the range of $1500-150000$ K for DA and $3250-150000$ K for DB type WDs, and for $\log g$ in the range of $7-9$ (cgs units) for both DA and DB WDs.} \citep{1995PASP..107.1047B} provide an estimate of the mass and cooling age for the WD as well as its synthetic photometry in all optical and infrared bands we studied. For a WD in the LAMOST catalog, we force $\log g = 7$ if it is below $7$, and force $\log g = 9$ if it exceeds $9$, since 7 and 9 are the lower and upper boundaries of the cooling models. $24$ of the parent sample have surface gravity out of this range.

\item[(ii)] SED templates of M/brown dwarfs. For late-type dwarf companions, empirical relations have been constructed between the spectral type and SDSS colors as well as absolute $J$-band magnitude \citep{2002AJ....123.3409H}. Alternatively,  We adopt the colors provided by \cite{2018ApJS..234....1B} involving Pan-STARRS, 2MASS, and {\it WISE} as a function of the spectral type, and interpolate to get colors of SDSS and UKIDSS filters. Since the absolute $J$-band magnitude is available only for dwarfs cooler than M6 type, we use the absolute magnitudes provided by \cite{2002AJ....123.3409H} for all of the M-, L- and T-types dwarfs for consistency. 

\item[(iii)] Radiation spectra of dust disks. We adopt \citet{2003ApJ...584L..91J} model to describe the dust disk emission, which assumes a geometrically flat, optically thick disk illuminated by the central WD and re-radiating at infrared. The temperature of each annulus in the disk is determined by the effective temperature ($T_{\rm eff}$) of and its distance ($R_{\rm ring}$) to the WD, i.e.,
\be
T_{\rm ring} \simeq \left( \frac{2}{3\pi}\right)^{1/4} \left( \frac{R_{\rm WD}}{R_{\rm ring}}\right)^{3/4} T_{\rm eff},
\label{eq-Tring}
\ee
where $R_{\rm WD}$ is the WD radius. Radiation from each annulus can be approximated as a blackbody spectrum, and its integration over the radius from the inner to outer edges of the disk gives the total disk flux:
\be
F_{\nu,\rm disk} &\simeq& 12\pi^{1/3} \frac{R^2_{\rm WD}\cos i}{D^2} \left(\frac{2k_B T_{\rm eff}}{3h\nu} \right)^{8/3}   \nonumber \\
&&\times \frac{h\nu^3}{c^2} \int^{x_{\rm out}}_{x_{\rm in}} \frac{x^{5/3}}{e^x -1} dx, 
\ee
where $i$ is the inclination of the disk, $D$ is the WD distance to the observer, $k_B$ is the Boltzmann's constant, $h$ is the Planck constant, c is the speed of light, and $x \equiv h\nu/(k_B T_{\rm ring})$. 
\end{itemize}

For the purpose of SED fitting, the magnitude in each photometric band is converted into the flux density according to the zero points provided by SVO Filter Profile Service \citep{2012ivoa.rept.1015R, 2020sea..confE.182R}. 
According to the measured $T_{\rm eff}$ and $\log g$ for each WD, we interpolate to obtain the absolute magnitude in various photometric bands according to WD cooling models \citep{1995PASP..107.1047B}. Then model flux is converted from the absolute magnitude by adopting {\it Gaia} distances of the WDs, and we add a free normalization to account for measurement errors of $\log g$ and distances. The cooling models are initially fit to the SDSS and Pan-STARRS fluxes of the WD by minimizing 
\be
\chi^2 = \sum_i \frac{(f_{i, \rm mod} - f_{i, \rm obs})^2}{\sigma^2_{i, \rm obs}},
\label{eq-chi2}
\ee
where $f_{i, \rm mod}$ is the model flux for the $i$th photometric band in optical, and $f_{i, \rm obs}$ and $\sigma_{i, \rm obs}$ are the observed flux and its $1\sigma$ error, respectively. 

Then an infrared excess with respect to the WD cooling model is searched with a significance expressed by the photometric and model uncertainties, i.e., 
\be
\chi_w = \frac{f_{i, \rm obs} - f_{i, \rm mod}}{\sqrt{\sigma^2_{i, \rm obs} + \sigma^2_{i, \rm mod}}} , 
\ee
where the model uncertainty is also considered and $\sigma_{i, \rm mod} =0.05 f_{i, \rm mod}$ is assumed \citep[e.g.,][]{2015MNRAS.449..574R, 2020ApJ...902..127X}.  The criterion we adopt for an excess is $\chi_w>3$ for {\it WISE} $W1$ band or longward, i.e., the excess is at a $>3\sigma$ significance level.  
Besides that, it should also be satisfied that the source image in this WISE band is unaffected by any known artifacts, i.e., the quality flag (cc$\_$flag) of the band is zero.

Once an excess is detected, we add an additional model component, i.e., a late-type dwarf companion or a dust disk.  
For the case of a dwarf companion, we adopt the WD distance provided by $Gaia$, and for each spectral type from M1 to T6, $\chi^2$ is calculated, and the lowest $\chi^2$ value corresponds to the best-fit spectral type of the companion. 

For the dust disk case, degeneracies exist between the inclination angle and inner disk boundary, especially for the case here that only one or two data points are available for excess modeling. Here we set the inclination and inner boundary of the disk as free parameters. 
The dust disk can extend to $\sim 100$ WD radii. Constraints on the outer disk boundary require photometric data at longer wavelengths, which is not the case here. We therefore take $R_{\rm out} = 80\ R_{\rm WD}$ and assume an initial face-on  inclination, i.e., $i=0 \degree$ \citep[e.g.,][]{2011ApJS..197...38D}.
The inner disk radii is capped by $50\ R_{\rm WD}$, and the lower boundary is set by the sublimation temperature ($T_{\rm sub}$) of dust grains, as indicated by Equation~(\ref{eq-Tring}). 
For silicate-dominated dust, $T_{\rm sub} \sim 1200$ K, and for pure carbon dust,  $T_{\rm sub} \sim 2100\ {\rm K}$  \citep{2011ApJS..197...38D}. We adopt a more relaxed value of $T_{\rm sub} = 3000$ K, as it was pointed out that the sublimation temperature of the dust could be much higher due to the high metal vapor pressure in the inner disk \citep{2012ApJ...760..123R}. 

The companion model is compared with the dust disk model via the reduced-$\chi^2$ ($\chi^2/d.o.f$) of SED fitting, and the model with lower $\chi^2$ value is preferred.

\section{Images}
\label{sec:image}

The infrared excess found in Section~\ref{sec:model} are mainly based on {\it WISE} observations, and due to its poor spatial resolution, a significant fraction of the excess could arise from background contamination and/or late-type stellar companions (possibly M dwarfs). LAMOST DR5 WD catalog already identified some WD+M dwarf binaries, while a fraction of M dwarf companions may be too faint to exhibit features in WD spectra and could be missed in the classification. The contaminants radiating in {\it WISE} bands are expected to radiate at a little bit shorter wavelengths since the spectrum is continuous. If it is not as cool as BDs, it should be detected in the $K$ band, and possibly $z$ band. 

\begin{figure*}
\centering
\includegraphics[width=0.9\textwidth]{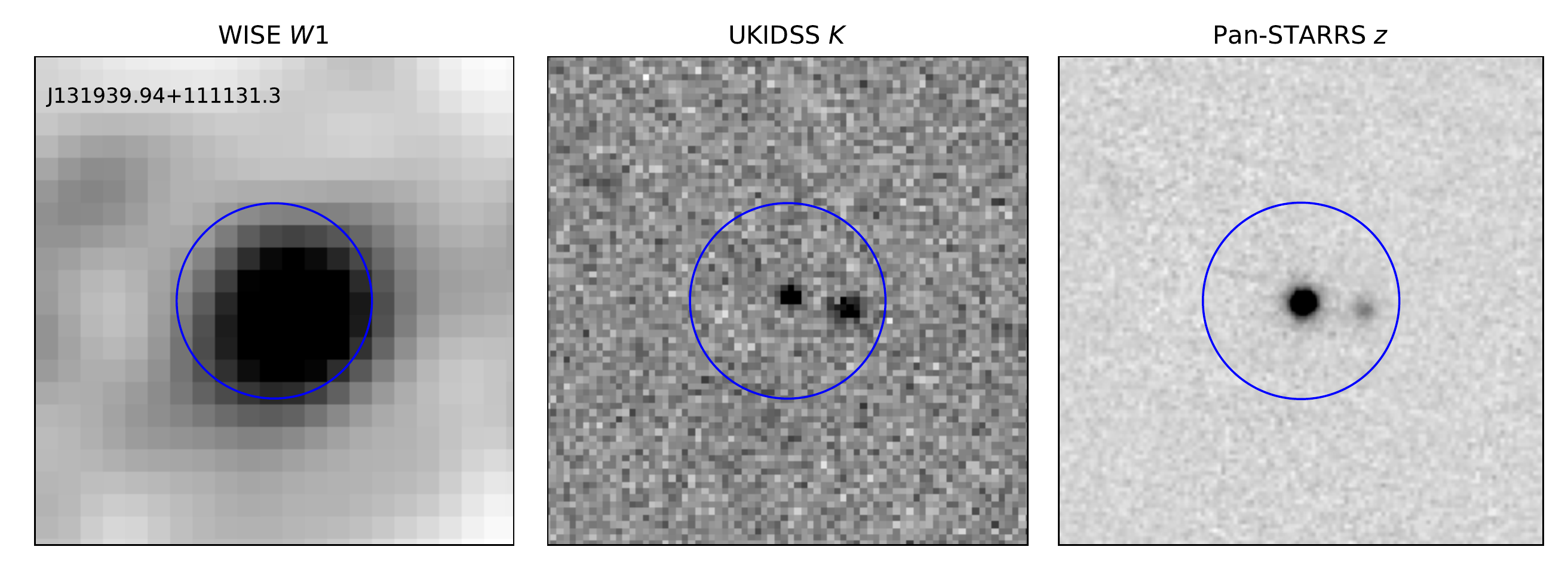}
\caption{A case of confused {\it WISE} $W1$ ($3.4\micron$; left panel) photometry classified as infrared excess according to SED model fitting but is removed from our candidate list due to contamination visible in the images. UKIDSS $K$- ($1.25\micron$; middle panel), and Pan-STARRS $z$-band (right panel) images are presented as comparison. The field of view is $30\arcsec \times 30\arcsec$ for each cutout, and the blue circle is centered at the LAMOST coordinate of the WD with a radius of 6\arcsec. As shown, Pan-STARRS can clearly distinguish the two point sources, though the contaminant is fainter compared to its UKIDSS image.}
\label{fig-img3}
\end{figure*}

UKIDSS image provides an effective tool to exclude background contamination owing to its superior spatial resolution. However, only a fraction of the WDs with infrared excess have UKIDSS observations. We notice that Pan-STARRS $z$ band data are also good in quality and may provide an alternative for contamination check, although the filters are even shorter in wavelength. As Figure~\ref{fig-img3} shows, the apparent two point-sources in the UKIDSS $K$-band image are blended in the {\it WISE} $W1$ image, and Pan-STARRS $z$-band can clearly resolve the two point sources. The circle with a radius of $6\arcsec$ can marginally enclose the central source in the ${\it WISE}$ image, which is also true for our other candidates with infrared excess. 
We checked the UKIDSS and Pan-STARRS images of all those WDs with infrared excess selected in Section~\ref{sec:model}, and those with multiple sources within a radius of $6\arcsec$ are excluded. While WD+BD binaries are not expected to show two point-sources in the $K$ or $z$ band, some of the WDs with an M dwarf companion and having images similar to Figure~\ref{fig-img3} may be excluded in this process. 

For the leftover candidates, we further compare their $WISE$ images and optical ones using the WISEView tool (http://byw.tools/wiseview), which is useful to look for possible contamination in the field. We remove those candidates with obvious contamination in the $WISE$ $W1$ image.

%%%%%%%%%%%%%%%%%%%%%%%%%%%%%%%%%%%%%%%%%%%%%%%%%

\section{Results and discussion}
\label{sec:result}
The SED fitting and image check result in a total of 50 candidates with infrared excess, including 7 candidate WD+M dwarf binaries, 31 candidate WD+BD binaries, and 12 candidate WD+dust disk systems. 37 out of the 50 candidates with infrared excess are new identifications, and 8 candidate dust disk systems are first reported as such systems. 

Our results rely upon the accuracy of $T_{\rm eff}$ measurement. An incorrect $T_{\rm eff}$ will result in an inaccurate measurement of WD photospheric radiation, possibly leading to a false or magnified excess in infrared. Moreover, our results may still suffer from background contamination. While UKIDSS and/or Pan-STARRS image check is able to remove contamination mainly radiating at shorter wavelength, those redder contaminants (e.g., red galaxies) may still be missed due to the limited resolution of $WISE$. These factors could lead to (very) high $\chi^2$ values in the SED fitting and should be treated with caution. Below we will list all selected candidates regardless of its $\chi^2$ value, and each of them waits for confirmation or refutation via follow-up observations. 

\subsection{WD+BD binary candidates}

Table~\ref{tab-1} lists the first part of WDs with candidate BD companions. Also listed are the 7 WDs with an M dwarf companion. WD+M dwarf binaries are not the focus of this paper, and our parent sample do not contain WDs spectroscopically classified as WD+M dwarf binaries by LAMOST, and the image check in Section~\ref{sec:image} may have excluded some of such systems, leaving only a few here. However, considering that the uncertainties of the spectral type of the companion could be several dex, we also list WD+M dwarf candidates here for reference. 

\begin{figure*}
\centering
\includegraphics[width=\textwidth]{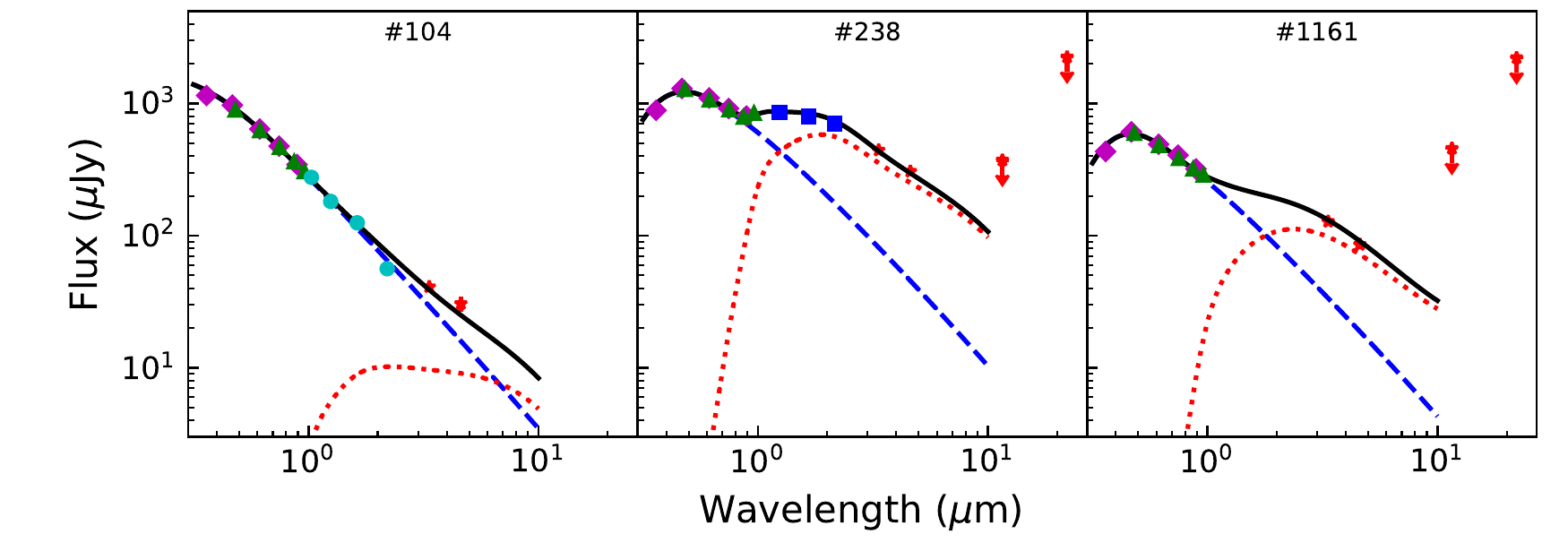}
\caption{Examples of SED fitting results for WD+BD candidates.  Colored symbols are photometric data from  SDSS (magenta diamonds), Pan-STARRS (green triangles), 2MASS (blue squares),  UKIDSS (cyan circles), and {\it WISE} (red stars). All data have been shown in the figure, and missing symbols in a panel means that there is no photometry in the corresponding survey (same for Figure~\ref{fig-disk}).  The flux in $W3$ and $W4$ bands given with upper limits are not used in the model fitting. The black solid line is the best-fitting model, which is the sum of WD photospheric radiation (blue dashed line) and BD contribution (red dotted line). The best-fit spectral types of the companion are L9$-$9.9, L0$-$0.9 and L5$-$5.9 BDs, respectively for the left, middle, and right panels. The y-axis errors of the data points are too small to be visible. }
\label{fig-binary}
\end{figure*}

The second part of the WD+BD candidates are listed in Table~\ref{tab-2}. These candidates have comparable reduced-$\chi^2$ resulting from the companion and the dust disk models. 
As examples, Figure~\ref{fig-binary} shows the SED fitting results for three WD+BD binaries. The best-fit spectral type of the BD companion is L9$-$9.9, L0$-$0.9 and L5$-$5.9, respectively for the left, middle, and right panels.

Two of previously known dust disk systems (WD 2328+107, ID: 470; and WD 0843+516, ID: 505) were misclassified as WD+BD binaries by our SED model fitting, mainly because $T_{\rm eff}$ given in the LAMOST catalog is higher than the best-fit value. An overestimated $T_{\rm eff}$ will shift blueward the model SED of the WD, creating spurious excess at shorter wavelengths, which mimics radiation from a late-type dwarf companion. When the best-fit $T_{\rm eff}$ is adopted, the resulting $\chi^2$ for the companion model and dust disk model are both reduced, and the dust disk model is now preferred.  This suggests that the identification of infrared excess and the best-fit model are sensitive to the accuracy of $T_{\rm eff}$ measurement. We have moved these two targets to the WD+dust disk classification, and for this reason we also consider the candidates in Table~\ref{tab-2} in the following statistical studies on dust disk properties.

\begin{deluxetable*}{cccccccccccc}
\tablewidth{0pt}
\tablecaption{WD+M/brown dwarf binary candidates  \label{tab-1}}
\tablehead{
\colhead{ID} & \colhead{R.A.} & \colhead{Dec}	& \colhead{G} & \colhead{Type}   & \colhead{T} & \colhead{$\log$ g} & \colhead{Mass}	& \colhead{Age} &\colhead{D} &\colhead{Spec. Type}  &\colhead{$\chi^2/d.o.f$} \\
 & \colhead{(deg)} & \colhead{(deg)} &  \colhead{(mag)} & & \colhead{(K)} & \colhead{($\rm cm\ s^{-2}$)} & \colhead{(M$_\odot$)} & \colhead{(Myr)} & \colhead{(pc)}    &         & 
}
\startdata
  162    &   236.821304    &    23.353057    &   16.5   &    DA    &  26345   &   7.84   &   0.56   &    14    &   145    &     T0-T0.9     &   36.2  \\
  238\tablenotemark{b}    &    1.948394     &    19.856837    &   16.2   &    DA    &  12123   &   8.26   &   0.77   &   538    &    87    &     L0-L0.9     &   2.7   \\
  343\tablenotemark{a}    &   137.593884    &    27.478865    &   17.2   &    DA    &  78945   &   7.77   &   0.66   &    1     &   602    &     M5-M5.9     &   7.2   \\
  769    &   186.738710    &    37.442958    &   17.4   &    DA    &  18042   &   7.67   &   0.46   &    60    &   240    &     L6-L6.9     &   14.6  \\
  1059\tablenotemark{a,b}   &   240.693580    &    30.654022    &   16.3   &    DA    &  71274   &   7.78   &   0.65   &    1     &   377    &     M6-M6.9     &   5.3   \\
  1080   &   241.631670    &    25.114203    &   17.6   &    DA    &  25836   &   7.92   &   0.59   &    16    &   266    &     L4-L4.9     &   15.2  \\
  1161   &    51.704255    &    18.465891    &   17.1   &    DA    &  13803   &   8.06   &   0.65   &   279    &   126    &     L5-L5.9     &   2.9   \\
  1278   &    65.135723    &    -1.405366    &   17.5   &    DA    &  16783   &   7.98   &   0.60   &   132    &   166    &     T1-T1.9     &   4.7   \\
  1345   &    5.956828     &    48.156015    &   16.7   &    DA    &  25314   &   8.06   &   0.67   &    25    &   170    &     T0-T0.9     &   8.2   \\
  1762\tablenotemark{b}   &   158.903457    &    -5.356090    &   16.9   &    DA    &  11209   &   8.62   &   1.00   &   1340   &   168    &     M7-M7.9     &   13.3  \\
  1898   &   202.352900    &    -0.945752    &   16.4   &    DA    &  23916   &   7.68   &   0.48   &    19    &   125    &     L9-L9.9     &   17.5  \\
  1923   &   106.356590    &    27.128318    &   17.8   &    DA    &   8635   &   9.00   &   1.20   &    71    &   378    &     M5-M5.9     &   36.9  \\
  2112   &   208.117270    &    9.177510     &   18.3   &    DA    &  36846   &   7.00   &   0.30   &    71    &   790    &     M5-M5.9     &   18.5  \\
  2225   &    27.470080    &    10.535010    &   17.5   &    DA    &  24274   &   7.82   &   0.54   &    20    &   250    &     L6-L6.9     &   27.4  \\
  2321   &   135.519260    &    26.137318    &   19.0   &    DA    &   9368   &   8.87   &   1.13   &   2530   &   550    &     M6-M6.9     &   53.8  \\
  2325   &    25.835801    &    36.478634    &   17.3   &    DA    &  34018   &   7.87   &   0.59   &    6     &   424    &     L0-L0.9     &   3.1   \\
  2423   &   114.262030    &    47.630261    &   17.8   &    DA    &  28584   &   7.59   &   0.46   &    11    &   444    &     M6-M6.9     &   79.2  \\
  2477\tablenotemark{a}   &   134.078957    &    16.184380    &   15.6   &    DB    &  31738   &   8.29   &   0.79   &    17    &    72    &     T1-T1.9     &   25.7  \\
  2534   &   163.697033    &    34.003756    &   16.8   &    DA    &  15163   &   7.59   &   0.41   &   104    &   169    &     L6-L6.9     &   4.1  \\
 \enddata
\tablecomments{Columns from left to right represent: (1)-(3) the unique ID \citep[same as the Group ID in ][]{2022MNRAS.509.2674G} and LAMOST coordinate; (4) G band magnitude given by $Gaia$ DR2; (5)-(9) spectral type, effective temperature, surface gravity, mass and cooling age of the WD given by LAMOST catalog; (10) {\it Gaia} distance of the WD; Columns (11)-(12): best-fit spectral type of the companion, and minimum reduced $\chi^2$. 
\tablenotetext{a}{The WD was reported by \cite{2011ApJS..197...38D} as candidate WD+M dwarf binaries.}
\tablenotetext{b}{The WD was identified as exhibiting {\it WISE} excess by \cite{2020ApJ...902..127X}.}
}
\end{deluxetable*}

\begin{deluxetable*}{cccccccccccccccc}
\tablewidth{0pt}
\tablecaption{WD+BD binary candidates with $\chi^2$ comparable to that results from the dust disk model \label{tab-2}}
\tablehead{
&&&&&&&&&&\multicolumn{2}{c}{Companion Model} & &\multicolumn{3}{c}{Disk Model} \\
\cline{11-12}
\cline{14-16}
\colhead{ID}  & \colhead{R.A.} & \colhead{Dec}	& \colhead{G} & \colhead{Type}   & \colhead{T} & \colhead{$\log$ g} & \colhead{Mass}	& \colhead{Age}  &\colhead{D} &\colhead{Spec. Type} &\colhead{$\chi^2/d.o.f$}  &&\colhead{$R_{\rm in}$}  & \colhead{$i$} & \colhead{$\chi^2/d.o.f$} \\
& \colhead{(deg)} & \colhead{(deg)} &  \colhead{(mag)} & & \colhead{(K)} & \colhead{($\rm cm\ s^{-2}$)} & \colhead{(M$_\odot$)} & \colhead{(Myr)} & \colhead{(pc)}     &   & && \colhead{(R$_{\rm WD}$)} & &
}
\startdata  
   63    &   178.028760    &    31.482751    &   17.9   &    DA    &   9205   &   7.95  &   0.57   &   706    &   112    &     T1-T1.9     &   0.7    &&   2.7    &   86.7   &   0.8   \\
  104\tablenotemark{b}    &   184.690380    &    26.808820    &   16.7   &    DA    &  26844   &   7.79  &   0.53   &    12    &   197    &     L9-L9.9     &   2.4    &&   38.0   &   88.4   &   2.5    \\
  204    &   246.978950    &    31.723409    &   18.5   &    DA    &  23372   &   7.70  &   0.48   &    20    &   370    &     L5-L5.9     &   4.9    &&   9.2    &   88.4   &   4.9    \\
  217    &   255.985680    &    28.223106    &   18.1   &    DA    &  18429   &   8.12  &   0.69   &   124    &   271    &     L6-L6.9     &   2.6    &&   6.7    &   87.4   &   3.1   \\ 
  272    &   166.892570    &    27.391105    &   18.4   &    DA    &  11273   &   8.25  &   0.76   &   643    &   256    &     L7-L7.9     &   2.7    &&   3.5    &   77.8   &   3.1   \\ 
  476    &   352.692900    &    0.398787     &   18.5   &    DA    &  19043   &   8.30  &   0.80   &   157    &   305    &     L4-L4.9     &   4.3    &&   7.0    &   80.5   &   4.3    \\
  534    &    58.001722    &    19.206418    &   17.7   &    DA    &  10635   &   8.69  &   1.03   &   1700   &   195    &     T6-T6.9     &   8.4    &&   13.9   &   0.0    &   8.4   \\ 
  737\tablenotemark{a,b}    &   161.749090    &    37.765775    &   16.9   &    DA    &  21536   &   8.08  &   0.67   &    60    &   180    &     L5-L5.9     &   3.7    &&   8.3    &   86.2   &   4.8    \\
  748    &   127.477780    &    48.874064    &   18.1   &    DA    &  19712   &   8.03  &   0.64   &    80    &   285    &     L6-L6.9     &   2.8    &&   7.3    &   87.5   &   3.2    \\
  848    &   218.117523    &    28.823069    &   17.2   &    DA    &  18745   &   7.89  &   0.56   &    73    &   181    &     L6-L6.9     &   5.2    &&   6.9    &   88.9   &   6.5   \\
  1163   &    50.610170    &    21.234753    &   17.8   &    DA    &  10704   &   8.37  &   0.84   &   892    &   134    &     T1-T1.9     &   0.8    &&   6.3    &   81.1   &   1.4    \\
  1633   &   185.237050    &    15.565191    &   17.1   &    DA    &  19635   &   7.95  &   0.59   &    66    &   174    &     L9-L9.9     &   3.3    &&   50.0   &   64.7   &   4.1   \\ 
  1714   &   190.110500    &    37.396327    &   17.3   &    DA    &  17971   &   7.93  &   0.58   &    93    &   181    &     T2-T2.9     &   3.1    &&  50.0   &   75.3   &   3.2   \\ 
  1794   &   138.567563    &    42.174152    &   17.2   &    DA    &  14000   &   8.06  &   0.65   &   268    &   134    &     L9-L9.9     &   8.6    &&   4.7    &   86.7   &   10.0   \\
  2020   &   155.261880    &    20.031298    &   17.4   &    DA    &  15170   &   8.04  &   0.64   &   205    &   155    &     L9-L9.9     &   1.2    &&   5.2    &   87.7   &   1.4   \\ 
  2070   &   206.417440    &    16.936465    &   17.7   &    DA    &  28830   &   7.95  &   0.62   &    10    &   349    &     L7-L7.9     &   2.1    &&   12.2   &   89.1   &   2.2    \\
  2351   &   136.410250    &    29.102879    &   17.1   &    DB    &  23448   &   7.89  &   0.55   &    27    &   184    &     L9-L9.9     &   5.8    &&   9.3    &   89.4   &   6.4    \\
  2410   &   126.987885    &    56.067105    &   17.7   &    DA    &  11888   &   7.79  &   0.49   &   295    &   217    &     L9-L9.9     &   2.6    &&   18.4   &   70.1   &   2.8    \\
  2810   &   140.478080    &    22.743359    &   17.6   &    DA    &   9347   &   7.97  &   0.58   &   695    &   104    &     T0-T0.9     &   0.7    &&   2.7    &   84.8   &   0.8    \\
\enddata
\tablecomments{Columns (1)-(10): similar to those in Table~\ref{tab-1}; (11)-(12): best-fit spectral type of the companion, and minimum reduced-$\chi^2$ for the companion model; (13)-(15): best-fit inner radius and inclination of the disk, and minimum reduced-$\chi^2$ for the dust disk model. 
\tablenotetext{a}{The WD was identified to have {\it WISE} excess by \cite{2020ApJ...902..127X}.
\tablenotetext{b}{The WD was confirmed to have {\it Spitzer} excess by \cite{2021ApJ...920..156L}.} 
}
}
\end{deluxetable*}

\subsection{WD+dust disk candidates}

Our WD+dust disk candidates are listed in Table~\ref{tab-3}. All these WDs have spectral types of either DA or DB as expected, since this is also the case for our parent sample. Although most of the dust disk candidates found by \cite{2011ApJS..197...38D} have similar stellar spectral types, WDs with known dust disks usually show metal absorption lines in WD spectra and are mostly classified as DAZ or DBZ. 
However, metal lines could be missed in the WD spectrum with a relatively low S/N. Three of the known WDs with a debris disk (WD 1018+410; ID: 282;  WD 0843+516, ID: 505; and Ton 345, ID: 923) classified as DAZ or DBZ in the literature \citep[see Table 1 of][]{2020ApJ...902..127X} are classified as DA or DB in the LAMOST catalog \citep{2022MNRAS.509.2674G}.  The effective temperatures of these WDs mostly locate at $\lesssim 25000$ K, consistent with the expectation that dust grains may not survive around a too hot WD. Indeed, those published WD+dust disk systems all have similar WD temperatures \citep[see][and references therein]{2020ApJ...902..127X}.  

\begin{figure*}
\centering
\includegraphics[width=\textwidth]{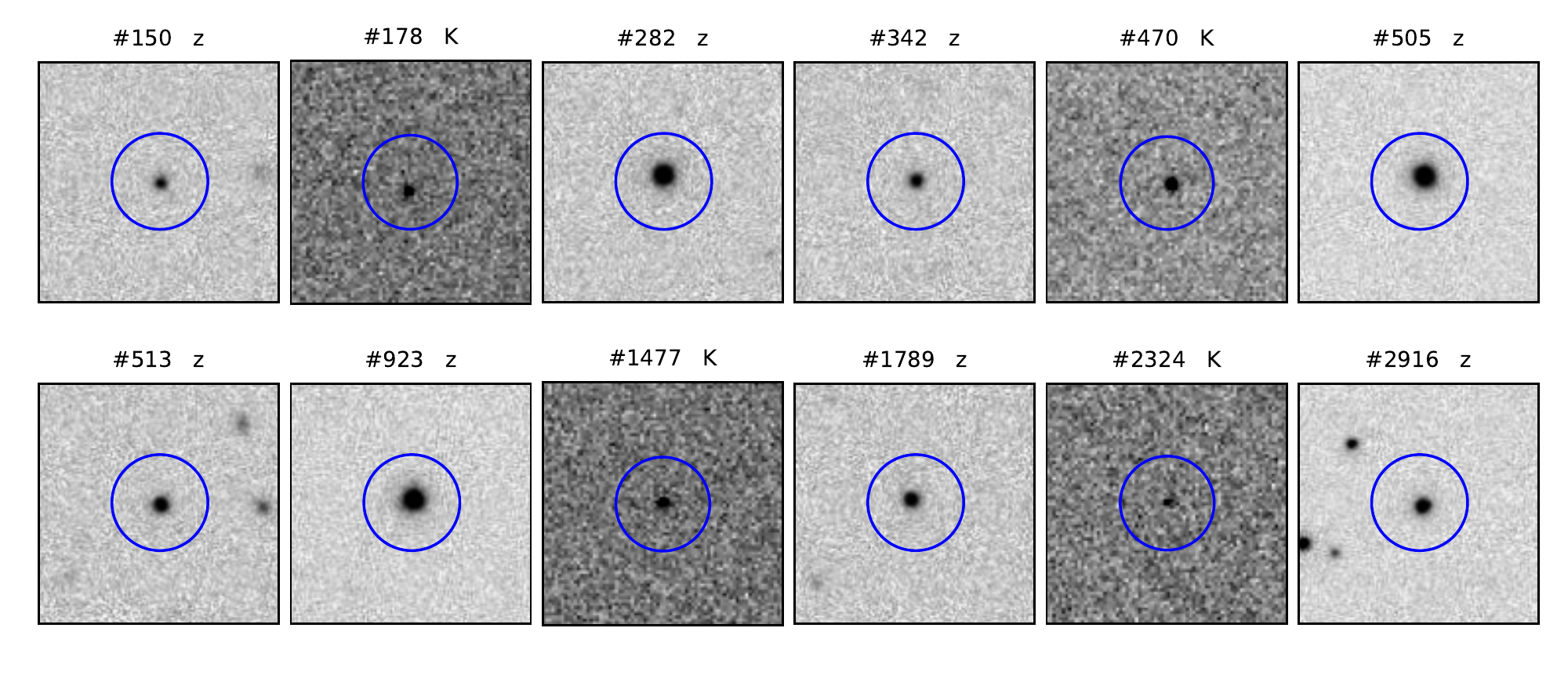}
\caption{UKIDSS $K$- or Pan-STARRS $z$-band images for WD+dust disk candidates, marked as `K' and `z' respectively on the top of each panel. Also labeled is the target ID listed in the first column of Table~\ref{tab-3}. The blue circle is centered at the LAMOST coordinate of the WD with a radius of 6\arcsec.}
\label{fig-img}
\end{figure*}

The cutout images of these candidate WD+dust disk systems are shown in Figure~\ref{fig-img}. The UKIDSS $K$-band image is our first choice, and if it is not available, the Pan-STARRS $z$-band image is shown. The blue circle centered at the LAMOST coordinate of the WD is to illustrate that there is no contamination within a radius of 6\arcsec.

\begin{deluxetable*}{ccccccccccccc}
\tablewidth{0pt}
\tablecaption{WD+dust disk candidates  \label{tab-3}}
\tablehead{
\colhead{ID}  & \colhead{R.A.} & \colhead{Dec}	& \colhead{G} & \colhead{Type}   & \colhead{T} & \colhead{$\log$ g} & \colhead{Mass}	& \colhead{Age}  &\colhead{D}  &\colhead{$R_{\rm in}$}  &\colhead{$i$} &\colhead{$\chi^2/d.o.f$} \\
& \colhead{(deg)} & \colhead{(deg)} &  \colhead{(mag)} & & \colhead{(K)} & \colhead{($\rm cm\ s^{-2}$)} & \colhead{(M$_\odot$)} & \colhead{(Myr)} & \colhead{(pc)}    &  \colhead{(R$_{\rm WD}$)}   & \colhead{(deg)} & 
}
\startdata
  150    &   189.683020    &    29.167623    &   18.8   &    DA    &  15000   &   8.59   &   0.98   &  520.0   &  227.1   &   8.1    &   65.8   &   0.9   \\
  178    &   158.522721    &    6.046522     &   16.3   &    DA    &  21360   &   7.97   &   0.61   &   46.0   &  137.6   &   11.4   &   89.4   &   3.4   \\
  282\tablenotemark{a}    &   155.481254    &    40.837750    &   16.4   &    DA    &  24431   &   8.21   &   0.76   &   49.2   &  135.7   &   9.8    &   87.9   &   8.2   \\
  342    &   130.685580    &    29.804366    &   18.5   &    DA    &  12705   &   8.24   &   0.76   &  460.0   &  284.3   &   4.8    &   77.7   &   1.6   \\
  470\tablenotemark{a}    &   352.673574    &    11.035077    &   15.6   &    DA    &  22675   &   7.84   &   0.55   &   26.7   &   98.0   &   8.8    &   89.6   &   12.2  \\
  505\tablenotemark{a}    &   131.759613    &    51.481411    &   16.1   &    DA    &  25259   &   7.96   &   0.61   &   19.6   &  139.0   &   10.2   &   87.7   &   5.9   \\
  513    &   142.004550    &    13.538704    &   17.9   &    DA    &  23788   &   7.88   &   0.57   &   22.9   &  286.7   &   9.4    &   85.6   &   1.0   \\
  923\tablenotemark{a}    &   131.413230    &    22.957784    &   15.9   &    DB    &  18806   &   8.15   &   0.68   &  125.0   &  105.2   &   6.9    &   82.2   &   1.3   \\
  1320   &   331.042745    &    2.558907     &   17.2   &    DA    &  16588   &   8.08   &   0.66   &  165.0   &  141.6   &   17.6   &   82.1   &   5.3   \\
  1477   &    28.959316    &    4.525195     &   16.8   &    DA    &  14000   &   7.94   &   0.58   &  224.0   &  125.3   &   4.7    &   88.7   &   1.9   \\
  1789   &   136.356200    &    39.727530    &   18.0   &    DA    &  15683   &   7.40   &   0.35   &   82.5   &  298.2   &   24.6   &   79.5   &   1.9   \\
  1953   &   164.532600    &    2.113474     &   16.9   &    DA    &  18667   &   8.05   &   0.65   &  103.0   &  163.8   &   46.8   &   73.2   &   1.8   \\
  2324\tablenotemark{b}   &   132.618090    &    27.332293    &   17.3   &    DA    &  20318   &   7.72   &   0.48   &   37.4   &  197.4   &   7.6    &   87.8   &   3.8   \\
  2916\tablenotemark{c}   &   224.171390    &    17.070909    &   17.0   &    DA    &  31375   &   7.93   &   0.61   &   7.6    &  245.8   &   13.6   &   88.7   &   1.2   \\
\enddata
\tablecomments{Columns (1)-(10): similar to those in Table~\ref{tab-1}; (11)-(13) best-fit inner radius and inclination of the dust disk, and minimum reduced-$\chi^2$.
\tablenotetext{a}{The WD is a previously known dust disk system.}
\tablenotetext{b}{The WD was reported as a candidate WD+M dwarf binary by \cite{2011ApJS..197...38D}.}
\tablenotetext{c}{The WD was identified with {\it WISE} excess by \cite{2020ApJ...902..127X}.}
}
\end{deluxetable*}

\begin{figure*}
\centering
\includegraphics[width=0.9\textwidth]{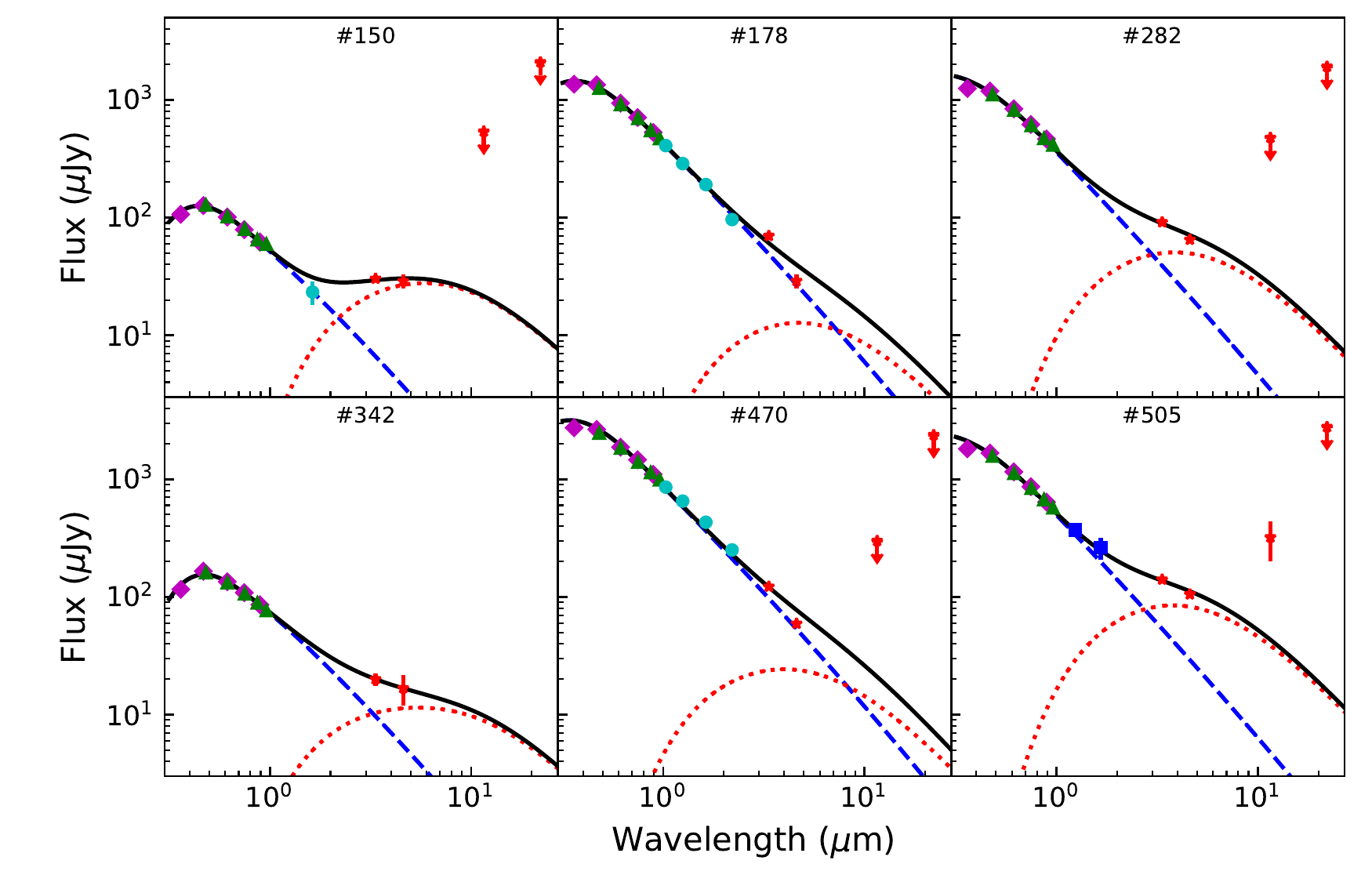}
\caption{SED fitting results for WD+disk candidates. Colored symbols are photometric data from SDSS (magenta diamonds),  Pan-STARRS (green triangles),  2MASS (blue squares),  UKIDSS (cyan circles), and {\it WISE} (red stars).  The black solid line is the best fitting model, which is the sum of contributions from WD photosphere (blue dashed line) and dust disk (red dotted line).}
\label{fig-disk}
\end{figure*}

The SED fitting results of the dust disk candidates are shown in Figure~\ref{fig-disk}. The disk inclinations of the candidates concentrate at $i>60\degree$ and the inner disk radii cover a broad range from several to tens of WD radii.   
As the disk is illuminated by the central WD, the disk temperature declines from the inner to outer radii.  The blue cutoff of the disk radiation is mainly set by the inner disk boundary and the declining rate at red end is determined by the outer disk boundary. The inclination can affect  the normalization, peak, and shape of the disk SED and is difficult to uniquely constrain due to degeneracies with the inner disk radius, especially for those targets with excess only in $W1$ band. Therefore, the best-fit parameters of those candidate dust disk systems should be treated as indicative if the dust disk is confirmed. 

\addtocounter{figure}{-1}
\begin{figure*}
\centering
\includegraphics[width=0.9\textwidth]{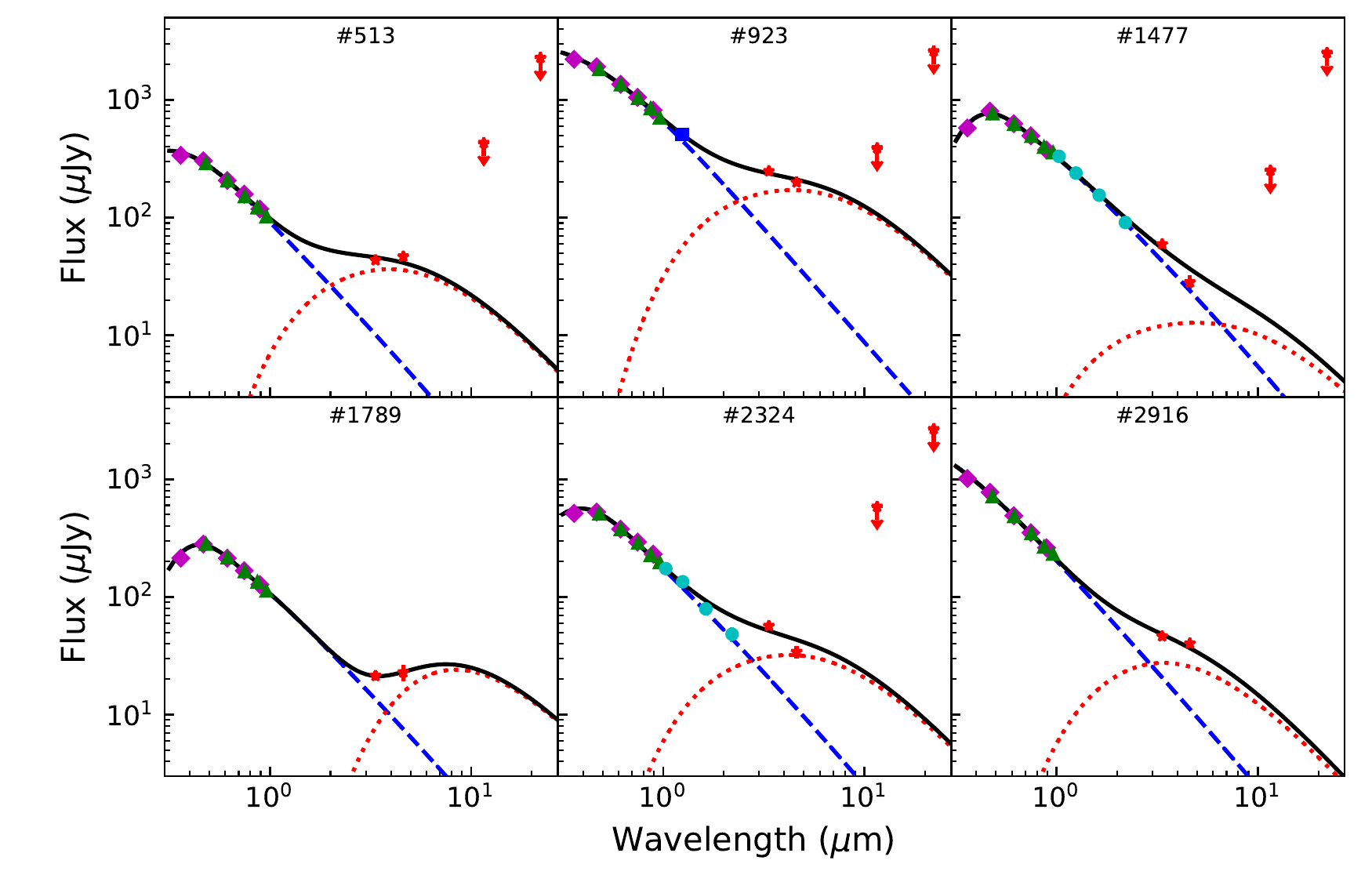}
\caption{--- $Continued$}
\end{figure*}

\begin{figure}
\centering
\includegraphics[width=0.49\textwidth]{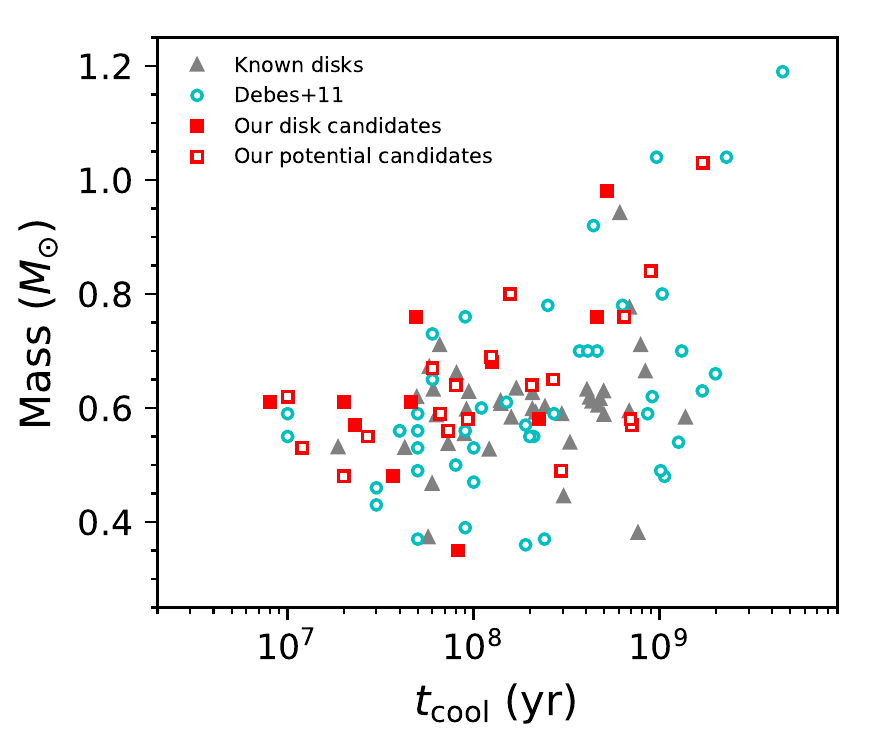}
\caption{Mass vs. cooling age of the WDs. Red filled squares are our dust disk candidates (in Table~\ref{tab-3}), and the red open squares are the potential dust disk candidates (in Table~\ref{tab-2}) with similar $\chi^2$ for the dust disk and the companion models.  WDs with known dust disk are shown as the gray triangles, and dust disk candidates identified by \citet{2011ApJS..197...38D} are marked as cyan circles. }
\label{fig-mt}
\end{figure}

Figure~\ref{fig-mt} shows mass vs. cooling age distribution of the WDs for the dust disk candidates.  Our candidates occupy similar parameter space to the WDs already known to host a dust disk. Including the targets with statistically similar fits between dust disk and companion models  in Table~\ref{tab-2} (red open squares) does not change the overall distribution.  

\begin{figure}
\centering
\includegraphics[width=0.49\textwidth]{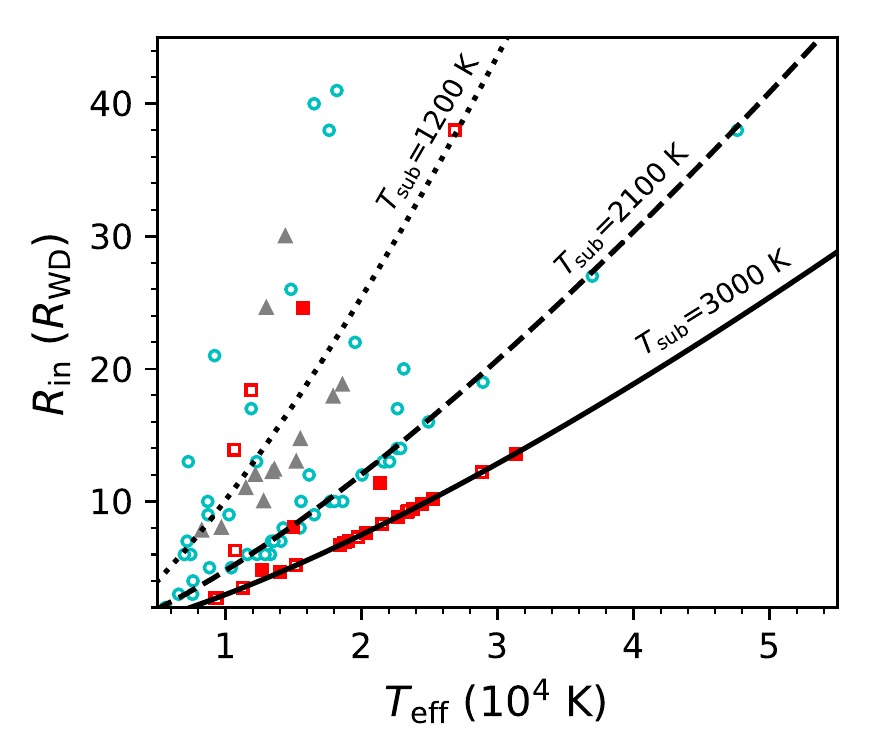}
\caption{Inner disk radius vs. the effective temperature of the WD for (candidate) dust disk systems. The legend is similar to Figure~\ref{fig-mt}. The dotted, dashed, and solid lines are the relations given by dust sublimation temperature of 1200 K, 2100 K and 3000 K, respectively.}
\label{fig-rt}
\end{figure}

In Figure~\ref{fig-rt}, we plot distribution of the inner disk radius ($R_{\rm in}$) vs. effective temperature of the WDs for our candidates. According to Equation~(\ref{eq-Tring}), we have $R_{\rm in} \propto (T_{\rm eff}/T_{\rm sub})^{4/3}$, where the sublimation temperature $T_{\rm sub}$ of dust grains defines a power-law relation between $R_{\rm in}$ and $T_{\rm eff}$ (black curves).  A significant fraction of our candidates locate at $T_{\rm sub} = 3000$ K, higher than the previously known dust disks and candidate dust disks identified by \citet{2011ApJS..197...38D}. This is simply caused by our settings of $T_{\rm sub} \le 3000$ K, which is to account for possible effects that may elevate sublimation temperature of dust grains \citep{2012ApJ...760..123R}. Beside that, the distribution of our candidates are in general consistent with the literature. 

The mass of the dust disk cannot be constrained if it is optically thick \citep[e.g.,][]{2003ApJ...584L..91J}. However, if the disk is optically thin, under the assumption of a modified blackbody radiation, the dust disk mass can be roughly estimated via the dust temperature $T_{\rm d}$ and the radiative flux $F_\nu$  of the disk at a specific wavelength $\lambda$, similar to the commonly used method to calculate the dust mass of protoplanetary disks and interstellar dust mass \citep[e.g.,][]{2009A&A...504..415S, 2019ApJ...874..141U}:
\be
M_{\rm dust} = \frac{F_\nu(\lambda) D^2}{\kappa(\lambda) B_\nu(T_{\rm d}, \lambda)},
\label{eq-Md}
\ee  
where $D$ is the WD distance, $\kappa (\lambda)$ is the dust absorption cross-section per unit mass at $\lambda$, and $B_\nu (T_d, \lambda)$ is the Planck function for temperature $T_d$ at $\lambda$. We take $\lambda = 2.2\ \micron$ and adopt $\kappa  (2.2\ \micron)=3800 \ {\rm cm^2 \ g^{-1}}$ \citep{2011ApJ...728..143S}. We have assumed multi-color blackbody radiation in the SED models described in Section~\ref{sec:model}. Here the dust temperature is assumed to be constant across the disk, and can be roughly estimated via the peak radiation of the dust SED according to Wien's displacement law. 
$F_\nu$ in Equation~(\ref{eq-Md}) is obtained from the model flux radiated by the dust disk, i.e., the red dotted lines in Figure~\ref{fig-disk}. 

\begin{figure}
\centering
\includegraphics[width=0.49\textwidth]{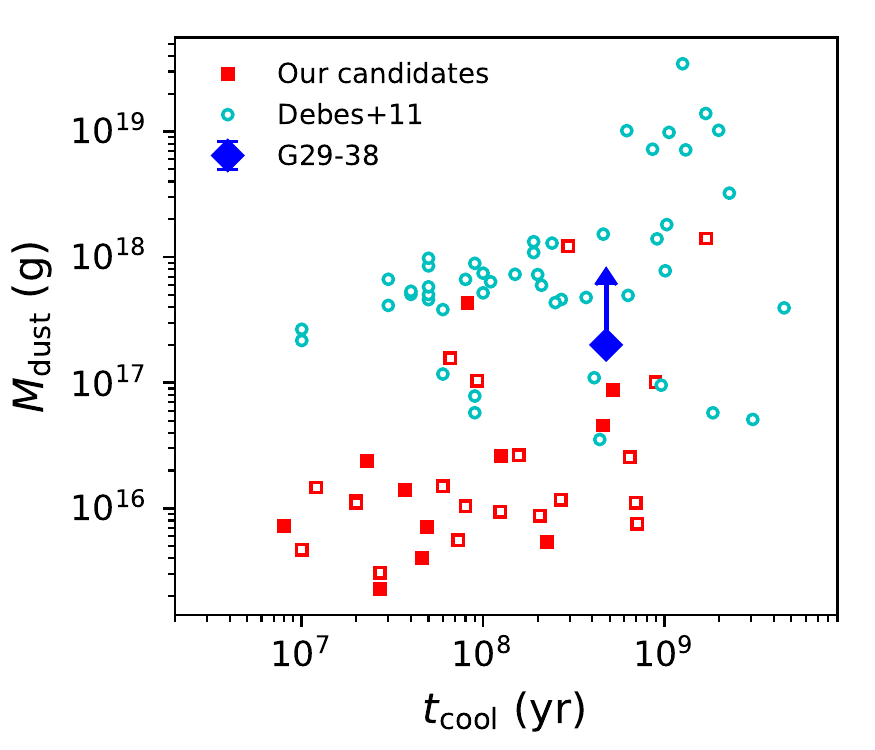}
\caption{Dust disk mass vs. WD cooling age for (candidate) dust disk systems. Filled (open) red squares are our candidates in Table~\ref{tab-3} (Table~\ref{tab-2}). Cyan circles are the dust disk candidates of \citet{2011ApJS..197...38D}. All the red squares and cyan circles represent lower limits for the dust mass.} The blue diamond show the lower limit of dust mass given by the literature (see the text).
\label{fig-Md}
\end{figure}

Figure~\ref{fig-Md} shows the estimated dust disk mass vs. WD cooling age for our candidate WD+dust disk systems. For most of the candidate systems, the dust disk mass locates at $\sim 10^{15} - 10^{18} \ {\rm g}$, much lower than the total mass of Saturn's rings \citep[$\sim 10^{22} \ {\rm g}$;][]{1993AREPS..21..487E}. 
As comparison, the dust disk of G29-38, a well studied dust disk system, was estimated to have a mass of $\gtrsim 2\times10^{17} - 10^{22}\ {\rm g}$ (the blue diamond in Figure~\ref{fig-Md}), which is sensitive to parameters such as dust grain size and mostly relies on the optically-thin disk assumption \citep{2003ApJ...584L..91J, 2005ApJ...635L.161R, 2008ApJ...674..431F, 2009ApJ...694..805F}.  For our candidates, the optically-thin disk assumption here will also underestimate the dust mass, and the red squares in Figure~\ref{fig-Md} should be treated as lower limits. Pearson correlation between the two parameters results in a coefficient of 0.57 (for the open and closed red squares), implying a positive correlation between the dust mass and WD cooling age. This might tentatively suggest disruption of planetesimals and settling of dust on the disk plane along the central WD's cooling life if such a relation is confirmed in the future.
Also shown is the dust disk candidates identified by \citet{2011ApJS..197...38D}, where the dust mass is similarly calculated according to Equation~(\ref{eq-Md}). The dust mass for their candidates are in general $1-2$ orders higher than ours. The main reason is that the disk inclination is fixed at $i=0\degree$ for most of their candidates.

\subsection{The occurrent rate}
\label{sec:rate}

We found 31 candidate WD+BD binaries from an input of $\sim3000$ WDs in the LAMOST catalog. However, the requirement of $W1$ band observations (in CatWISE or AllWISE), $T_{\rm eff}$ measurements and {\it Gaia} distances left only 846 targets for excess search via SED model fitting. The event rate of WD+BD binaries is therefore $31/846\approx3.7\%$, about twice higher than value ($0.5-2$ percent) reported in the literature \citep[e.g.,][]{2011MNRAS.416.2768S, 2011MNRAS.417.1210G}. However, our frequency should be treated as an upper limit. Two of the previously known dust disk systems were originally identified as WD+BD binaries by us, with one listed in Table~\ref{tab-1} and the other in Table~\ref{tab-2}. We expect that a significant fraction of the candidate WD+BD binaries, especially those listed in Table~\ref{tab-2}, could be dust disk systems if the excess is real. Follow-up observations such as radial velocity measurement of WDs \citep[e.g.,][]{2006Natur.442..543M, 2013MNRAS.429.3492S} or high resolution infrared imaging \citep[e.g.,][]{2019MNRAS.487..133W} may confirm or refute such systems.

The twelve WD+dust disk candidates we identified convert to a frequency of $12/846\approx 1.4\%$ for such systems, consistent with $1-4$ percent reported in the literature \citep[e.g.,][]{2015MNRAS.449..574R}. If we include the 19 potential dust disk candidates in Table~\ref{tab-2} with comparable $\chi^2$ for dust disk and companion models, the frequency of dust disks will be $3.7\%$. We reaffirm that for some of them, the infrared excess could still be spurious if $T_{\rm eff}$ measurement is inaccurate or could be caused by background contamination.  Moreover, the final 846 sample left for SED model fitting may suffer from selection biases, since only WDs of DA or DB types with relatively high S/N spectra have $T_{\rm eff}$ measurements in the LAMOST catalog. 

 \subsection{Comparison with the literature} 
 
 The 50 candidates we found with infrared excess include 4 WDs with previously known dust disks, 5 targets already reported by \citet{2020ApJ...902..127X} as exhibiting unWISE excess, and 7 sources studied by \citet{2020ApJ...902..127X} yet no excess found by them. Our different results could arise from several factors, e.g., (i) the effective temperature of WDs is obtained by profile-fitting of Balmer absorption lines in LAMOST WD catalog, which could be different from that derived from {\it Gaia}'s auto fit;  (ii) we set the excess criterion as exceeding the WD photospheric radiation at $>3\sigma$ level, while they adopt a more significant $>5\sigma$ level; (iii) the CatWISE and/or AllWISE photometric data we used are not exactly the same as the unWISE data they adopted. 2 out of the 5 targets commonly identified as exhibiting {\it WISE} excess by us were confirmed to have {\it Spitzer} excess \citep{2021ApJ...920..156L}. 
 
Besides one WD with known dust disk, 4 of our candidates with infrared excess were also identified by \citet{2011ApJS..197...38D}. We have 2 common candidate WD+M dwarf binaries. 2 candidates classified as WD+M dwarf binaries by them are WD+BD and WD+dust disk systems respectively in our classification. One of these two targets were also studied by \citet{2020ApJ...902..127X}, who identified it as with no excess, neither in magnitudes nor in colors. An insignificant infrared excess may lead to different classifications by \citet{2011ApJS..197...38D} and us. 
 
\section{Conclusions}
\label{sec:conclusion}

We search for infrared excess around 3064 WDs recently released by LAMOST via SED model fitting of 846 targets with $WISE$ detections, effective temperature and distance measurement. The WD's effective temperature provided by LAMOST catalog is adopted in the SED fitting, and thus the identification of candidates with infrared excess relies on the accuracy of effective temperature measurement. Possible contamination have been preliminary excluded by visual check of optical and infrared images of these candidates. We identified 50 candidates with infrared excess, including 7 candidate WD+M dwarf binaries, 31 candidate WD+BD binaries, and 12 candidate WD+dust disk systems.  The frequencies of WD+BD binaries and WD+dust disk systems are estimated to be $\lesssim 3.7\%$ and $\sim1.4\%$, respectively. 8 of these WD+dust disk candidates are first reported as such systems. Our dust disk candidates occupy similar parameter space to previously known dust disk systems in the WD mass vs. cooling age plot.  The SED model fitting suggests that the temperatures of the inner dust disks locate at $\lesssim1200-3000$ K, generally consistent with that reported in the literature. The dust masses are constrained to have a lower limit of $\sim 10^{15} - 10^{18} {\rm g}$ under the assumption of optically-thin dust disks.  Positive correlations are found between the dust mass and WD cooling age. We caution that all these candidates with infrared excess require  follow-up infrared imaging or infrared spectroscopy for confirmation.

\acknowledgements
%\begin{acknowledgements}
%{\it Acknowledgements.} 
We thank the anonymous referee for his/her very constructive and helpful comments that have greatly improved the paper. We thank Dr. Siyi Xu for her quick and patient response to our email. X.Z. acknowledges her debt to Dr. Tianwen Cao, Liyuan Lu, Dr. Mouyuan Sun, and Qingzheng Yu for various helps in the aspects that she is not familiar with. This work is supported by the National Key Program for Science and Technology Research and Development under No. 2017YFA0402600, and  the National Natural Science Foundation of China under Nos. 11890692, 12003024, 12033004, 12103041, 12133008, 12203006, and 12221003. We acknowledge the science research grants from the China Manned Space Project with No. CMS-CSST-2021-A04. L.W. and X.Z. contribute equally to this work and are co-first authors.
%\end{acknowledgements}

\vspace{3mm}

\software{Astropy \citep{2013A&A...558A..33A, 2018AJ....156..123A}, 
	Matplotlib \citep{4160265}, 
	SciPy \citep{2020SciPy-NMeth},  
	Pandas \citep{2022zndo...3509134R}.
	}

%\clearpage
\bibliography{main}{}
\bibliographystyle{aasjournal}
\end{CJK*}
\end{document}